\documentclass[11pt]{article}
\usepackage{graphicx}
\usepackage{cite}
\usepackage{times}
\usepackage{amsmath}

\newcommand\pubnumber{TTP-19-002}
\newcommand\pubdate{7 January 2019}
\newcommand{\nohref}[1]{}

\def\napoli{Institut f\"ur Theoretische Teilchenphysik (TTP)\\
Karlsruhe Institute of Technology, 76131 Karlsruhe, Germany}

\def\Title#1{\begin{center} {\Large #1 } \end{center}}
\def\Author#1{\begin{center}{ \sc #1} \end{center}}
\def\Address#1{\begin{center}{ \it #1} \end{center}}

\newcommand\pubblock{\rightline{\begin{tabular}{l} \pubnumber\\
         \pubdate  \end{tabular}}}
\newenvironment{Abstract}{\begin{quotation}  }{\end{quotation}}
\newenvironment{Presented}{\begin{quotation} \begin{center} 
             Presented at\end{center}\bigskip 
      \begin{center}\begin{large}}{\end{large}\end{center} \end{quotation}}
\def\Acknowledgements{\bigskip  \bigskip \begin{center} \begin{large}
             \bf Acknowledgements \end{large}\end{center}}

\newcommand{\lqcd}{\Lambda_{\textit{\scriptsize{QCD}}}}

\newcommand{\beq}[1]{$$ {#1} $$}
\newcommand{\beqin}[1]{$ {#1} $}
\newcommand{\bfr}{\begin{frame}}
\newcommand{\efr}{\end{frame}}

\newcommand{\bo}{\raise-1mm\hbox{\Large$\Box$}} 

\newcommand{\bra}[1]{\langle #1|} \newcommand{\ket}[1]{|#1\rangle}

\newcommand{\eb}{\begin{equation}} \newcommand{\ee}{\end{equation}}

 \newcommand{\lt}{\left}
  \newcommand{\rt}{\right}

\newcommand{\gev}{\,\mbox{GeV}}

\newcommand{\Bbar}{\bar B}

\newcommand{\bbs}{${B_s}\!-\!\ov{{\!B}}{}_{s}\,$}
\newcommand{\bbms}{${B_s}\!-\!\ov{{\!B}}{}_{s}\,$\ mixing}

\newcommand{\dm}{\ensuremath{\Delta m}}
\newcommand{\dg}{\ensuremath{\Delta \Gamma}}

\newcommand{\bea}{\begin{eqnarray}}
\newcommand{\eea}{\end{eqnarray}}

\newcommand{\nn}{\nonumber \\}

\newcommand{\ov}{\overline}

\newcommand{\eq}[1]{Eq.~(\ref{#1})}

\begin{document}
\begin{titlepage}
\pubblock

\vfill
\Title{The width difference in the \bbs system: towards NNLO}
\vfill
\Author{Ulrich Nierste}
\Address{\napoli}
\vfill
\begin{Abstract}
  The width difference $\dg$ among the two mass eigenstates of the \bbs
  system is measured with a precision of 7\%. The theory
  prediction has a larger uncertainty which mainly stems from unknown
  perturbative higher-order QCD corrections. I discuss the subset of
  next-to-next-to-leading order diagrams proportional to
  $\alpha_s^2\, N_f$, where $N_f=5$ is the number of quark flavours.
  The results are published in \cite{Asatrian:2017qaz}.
\end{Abstract}
\vfill
\begin{Presented}
\emph{10th International Workshop on the CKM Unitarity
Triangle (CKM 2018)},\\ Heidelberg, Germany, September 17-21, 2018
\end{Presented}
\vfill
\end{titlepage}
\def\thefootnote{\fnsymbol{footnote}}
\setcounter{footnote}{0}

\section{\boldmath\bbms\ and \dg}
\begin{figure}[tb]
\centering
\includegraphics[height=3cm]{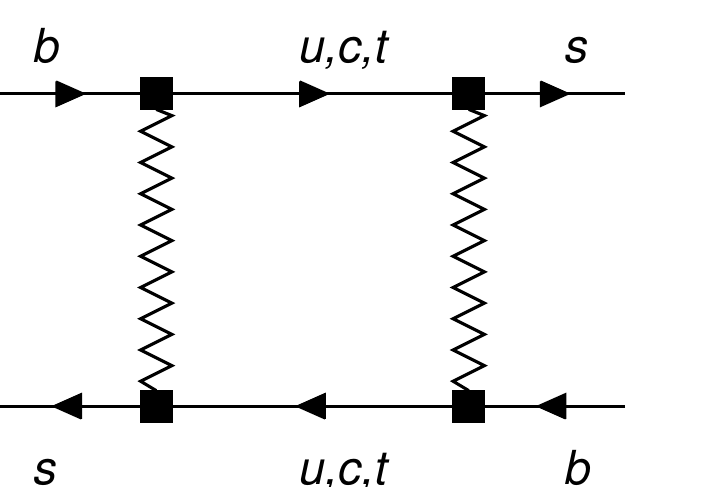}
\caption{Box diagram describing \bbms. A second diagram is obtained by
a 90$^\circ$ rotation.}
\label{fig:box}
~\\[-2mm]\hrule
\end{figure}
The box diagram of Fig.~\ref{fig:box} describes \bbms, which is a
transition changing the beauty quantum number $B$ by two units. 
As a consequence of \bbms, the flavour eigenstates $B_s$ and $\Bbar_s$
are not equal to the mass eigenstates $B_H$ and $B_L$ which obey
simple exponential decay laws. Denoting masses and decay widths of
$B_{H,L}$ by $M_{H,L}$ and $\Gamma_{H,L}$ (with the subscripts denoting
``heavy'' and ``light''), the mixing problem involves five
observables:
\begin{align}
M &= \frac{M_L+M_H}{2},  &
\Gamma &= \frac{\Gamma_L+\Gamma_H}{2}, & 
\dm &= M_H-M_L, & \dg &= \Gamma_L-\Gamma_H, &
\label{eq:5q}
\end{align}
and the CP asymmetry in flavour-specific decays, $a_{\rm fs}$, which
quantifies CP violation in mixing and is typically measured in
semileptonic decays. The mass difference
$\dm = (17.757 \pm 0.021)\, \mbox{ps}^{-1}$ \cite{hfag} has been
determined very precisely from the \bbs\ oscillation frequency
\cite{Abulencia:2006ze,Aaij:2013mpa}. The experimental value of the
width difference \cite{hfag},
\begin{align}
\dg^{\rm exp}  &= (0.088 \pm 0.006)\, \mbox{ps}^{-1},
\label{eq:dgexp}
\end{align}
is an average of measurements by LHCb \cite{lhcb,lhcb2}, ATLAS
\cite{Aad:2016tdj}, CMS \cite{Khachatryan:2015nza}, and
CDF \cite{Aaltonen:2012ie}. 

\begin{figure}[tb]
\centering
\includegraphics[height=2.7cm]{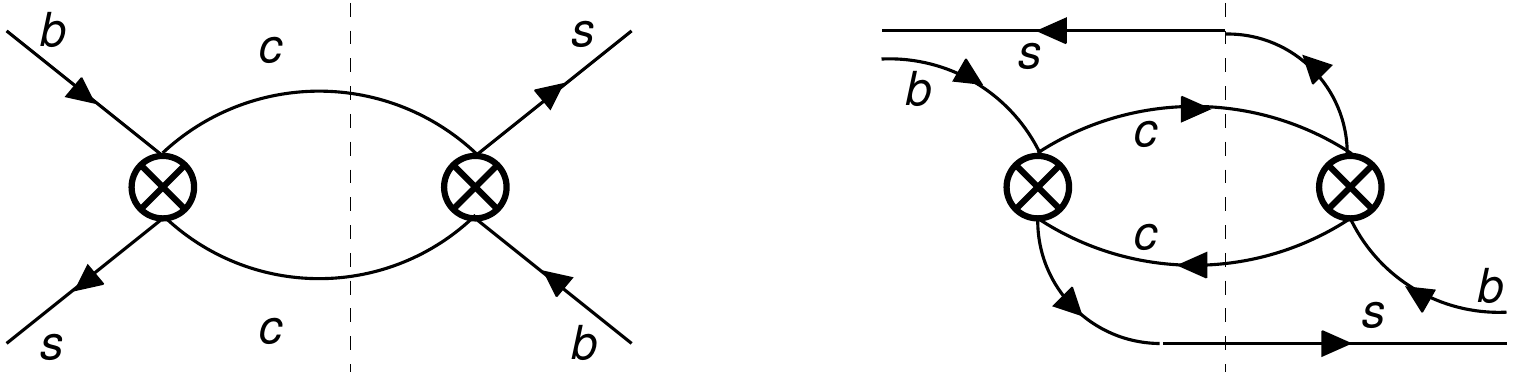}\hfill
\includegraphics[height=2.7cm]{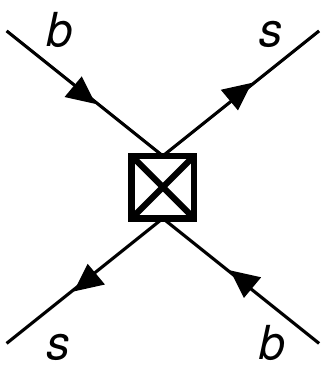}
\caption{Left and middle: LO diagrams for $\dg$ corresponding to the box
  diagrams of Fig.~\ref{fig:box}. The crosses represent the operators
  $O_1$ or $O_2$ from $H^{\Delta B=1}_{eff}$ in \eq{eq:heff}. Right:
  Effective $|\Delta B|=2$ operator.}
\label{fig:eff}
~\\[-2mm]\hrule
\end{figure}

$\dg$ is calculated from the absorptive part of the box diagram in
Fig.~\ref{fig:box}, which is the piece of this diagram involving the
imaginary part of the loop integral. Only the contributions with light
$u$ and $c$ quarks contribute to $\dg$. In order to include
strong-interaction effects one exploits that the bottom mass $m_b$ is
much larger than the fundamental scale of QCD, $\lqcd$, and employs an
operator product expansion, the \emph{heavy quark expansion (HQE)}
\cite{hqe,hqe2,hqe3,hqe4}. This procedure results in a systematic
expansion of $\dg$ in powers of $\lqcd/m_b\approx 0.1$ and
$\alpha_s(m_b)\approx 0.2$. At the energy scale $m_b$, relevant for
$B_s$ decays, $W$ exchange can be described by point-like
interactions. The corresponding effective $|\Delta B|=1$ hamiltonian for
the $b\to s$ transitions of our interest reads
\begin{eqnarray}
\label{Heff} H^{\Delta
B=1}_{eff}=\frac{G_F}{\sqrt{2}}V^*_{cs}V_{cb}\left\{
\sum^6_{i=1}C_iO_i + C_{8}O_{8}\right\} \; +\; \mbox{H.c.},
\label{eq:heff}
\end{eqnarray}
with
\begin{eqnarray}
  O_1&=&(\bar{s}_ic_j)_{V-A}\;(\bar{c}_jb_i)_{V-A}, \qquad
O_2\;=\;(\bar{s}_ic_i)_{V-A}\;(\bar{c}_jb_j)_{V-A}, \nn
O_{8}&=&\frac{g_s}{8\pi^2}m_b\bar{s}_i\sigma^{\mu\nu}(1-\gamma_5)T_{ij}^a
b_jG^a_{\mu\nu}.\label{OperBasis}
\end{eqnarray}
The numerically less important four-quark penguin operators $Q_{3-6}$
are not shown. $G_F$ is the Fermi constant, $i,j$ are colour indices,
$V\pm A= \gamma_{\mu}(1\pm \gamma_5)$, and $V_{cs}$ and $V_{cb}$ are
elements of the Cabibbo-Kobayashi-Maskawa (CKM) matrix. Note that the
doubly Cabibbo-suppressed contributions with $u$ quarks have been
neglected. The chromagnetic operator $O_8$ encodes a $b$-$s$-gluon and a
$b$-$s$-gluon-gluon coupling.  The Wilson coefficients $C_j$ in
\eq{eq:heff} comprise the short-distance QCD effects of the energy scale
of the $W$ mass and above. They are known to next-to-next-to-leading
order (NNLO) of QCD \cite{Gorbahn:2004my,Gorbahn:2005sa}. The
leading-order (LO) contribution to $\dg$ in both expansion parameters
$\alpha_s(m_b)$ and $\lqcd/m_b$ is shown in the left two diagrams of
Fig.~\ref{fig:eff}.
$\dg$ can be understood to come from the interference of all $B_s \to f$
and $\Bbar_s \to f$ decays, where $f$ is any final state common to $B_s$
and $\Bbar_s$ decays. The Cabibbo-favoured contribution to $\dg$ stems
from $b \to c \bar c s$ decays. The final state is indicated by the
dashed line in Fig.~\ref{fig:eff}. The results of the left and middle
diagrams in Fig.~\ref{fig:eff} determine the LO coefficients of the
effective $|\Delta B|=2$ operators, depicted at right in
Fig.~\ref{fig:eff}. At leading order of $\lqcd/m_b$ (``leading power'')
one needs two such operators:
\beq{ Q=(\bar{s}_ib_i)_{V-A}\;(\bar{s}_jb_j)_{V-A}, \qquad\quad
  \tilde{Q}_S=(\bar{s}_ib_j)_{S-P}\;(\bar{s}_jb_i)_{S-P}.}
Here $S- P= 1 - \gamma_5$.  Higher-order QCD corrections are
calculated from diagrams involving gluons added to the diagrams of
Fig.~\ref{fig:eff}, penguin diagrams, and diagrams involving $Q_8$.
Finally, non-perturbative QCD effects are contained in the
$|\Delta B|=2$ matrix elements:
\begin{align} \bra{B_s} Q (\mu_2) \ket{\ov B_s}  &=\;
   \frac{8}{3} M^2_{B_s}\, f^2_{B_s} B(\mu_2)\nn
\bra{B_s} \widetilde Q_S (\mu_2)\ket{\ov B_s} 
& =\; \frac{1}{3}  M^2_{B_s}\,
f^2_{B_s} \widetilde B_S^\prime (\mu_2). \label{eq:me}
\end{align}
Here $M_{B_s}$ and $f_{B_s}$ are mass and decay constant of the
$B_s$ meson, respectively, and $ \mu_2={\cal O}(m_b)$ is the
renormalisation scale at which the matrix elements are calculated. The
dimensionless quantities $B(\mu_2)$ and $ \widetilde B_S^\prime (\mu_2)$
parametrise the matrix elements.  The leading-power result can be
written as
\begin{align}
  \dg & =\, \frac{G_F^2
    m_b^2}{12 \pi\, M_{B_s}} \, |V_{cs}^* V_{cb}|^2 \lt| \, G^\prime \,
  \bra{B_s } Q \ket{\Bbar_s} \; + \; \widetilde G_S \,\bra{B_s}
  \widetilde Q_S \ket{\Bbar_s} \rt|
\label{eq:dghqe}
\end{align}  
with perturbative coefficients \beqin{G^\prime},\beqin{\widetilde
  G_S}. These coefficients are bilinear in the $C_j$'s of
$H^{\Delta B=1}_{eff}$ and are known to next-to-leading order (NLO) in
$\alpha_s(m_b)$
\cite{NiersteNLO,Ciuchini:2003ww,Beneke:2003az,NiersteNLONB}.  The
Wilson coefficients $C_j$ depend on an unphysical renormalisation scale
$\mu_1={\cal O}(m_b)$. The dependence of
\beqin{G^\prime},\beqin{\widetilde G_S} on $\mu_1$ diminishes
order-by-order in $\alpha_s$ and serves as an estimate of the accuracy
of the perturbative calculation. Also the dependence on the chosen
renormalisation scheme decreases with higher orders of $\alpha_s$.  For
instance, we can trade the pole mass $m_b$ in \eq{eq:dghqe} for the
$\ov{\rm MS}$ mass \beqin{\bar m_b} and replace e.g.\ $\widetilde G_S$
by
\beq{\widetilde G_S^{\overline{\rm MS}} \equiv \frac{m_b^{\rm pole\,
      2}}{\bar m_b^2} \widetilde G_S, }
expanded in \beqin{\alpha_s} to the order in which $\dg$ is calculated.
Corrections to $\dg$ of order $\lqcd/m_b$ involve additional operators
and have been calculated in Ref.~\cite{HNSBsBsbar}.

Including all known corrections one has
\begin{align}
  \dg & = \lt( 0.091 \pm 0.020_{\rm scale} \pm 0.006_{B,\widetilde
B_S}
       \pm  0.017_{\lqcd/m_b} \rt) \, \gev
\qquad \mbox{(pole)} \nn
\dg &= \lt( 0.104 \pm 0.008_{\rm scale} \pm
   0.007_{B,\widetilde B_S}
   \pm 0.015_{\lqcd/m_b} \rt)\, \gev
      \qquad \mbox{($\ov{\rm MS}$)} \label{eq:nlo}
\end{align}      
These numbers are found from the expressions in Ref.~\cite{NiersteNLONB}
with present-day lattice-QCD results for the matrix elements in
\eq{eq:me} taken from Ref.~\cite{Bazavov:2016nty}. The uncertainties
from different sources are indicated in \eq{eq:nlo}. The size of the
missing $\alpha_s^2$ corrections to the diagrams in Fig.~\ref{fig:eff}
can be estimated from the $\mu_1$-dependence, denoted with ``scale'', or
from the difference between the central values in the two schemes.  This
perturbative error is larger than the uncertainty stemming from the
lattice-QCD calculation denoted with ``$B,\widetilde B_S$'' and also
exceeds the experimental error in \eq{eq:dgexp}. Also the last error
related to the power corrections originates mostly from the unknown NLO
corrections to coefficients of the subleading-power operators. The
matrix elements of these subleading operators have been estimated with
QCD sum rules \cite{Mannel:2007am} and lattice-QCD calculations are
making progress \cite{Davies:2017jbi}. Thus perturbative uncertainties
are dominant and call for the calculation of the NNLO corrections to the
leading power contribution. Also NLO corrections to the $\lqcd/m_b$
piece are needed. The phenomenology of \dg\ within and beyond the
Standard Model is dicussed in
Refs.~\cite{Dunietz:2000cr,NiersteNLONB,Jager:2017gal}.

\section{Towards NNLO}
\begin{figure}[tb]
\centering
\includegraphics[width=0.85\textwidth, viewport=67 255 570 755]{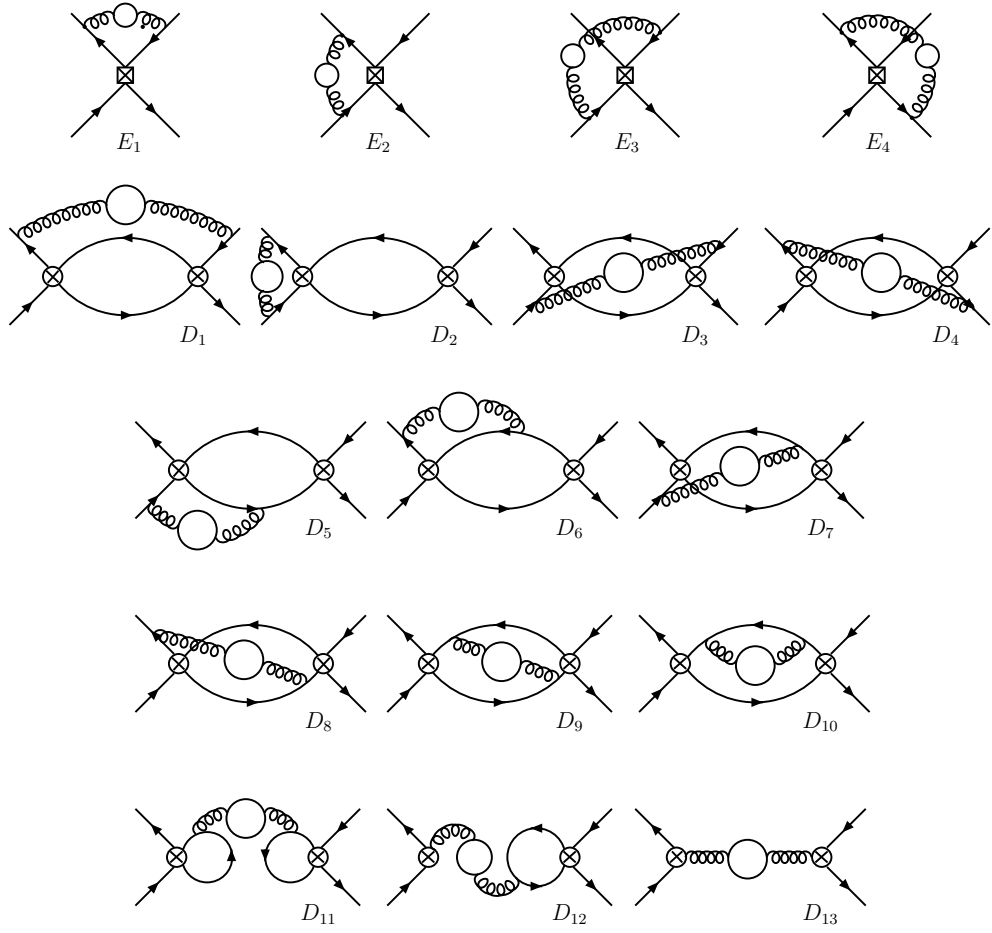}
\caption{Diagrams with $O_{1,2}$ or $O_8$ contributing to the
  ${\cal O}(\alpha_s^2 N_f)$ corrections to \dg.}
\label{fig:diags}
~\\[-2mm]\hrule
\end{figure}

The NNLO calculation involves three-loop diagrams with loop integrals
depending on one external momentum $p$ with $p^2=m_b^2$, i.e.\ these are
propagator-type integrals. $m_b$ and the charm mass $m_c$ appear on
internal lines. The calculation in Ref.~\cite{Asatrian:2017qaz} has
addressed the subset of diagrams with a closed quark loop, shown in
Fig.~\ref{fig:diags}.
These diagrams are a gauge-invariant subset of all NNLO diagrams and
grow with the number $N_f$ of active quark flavours. (In $b$ decays one
has $N_f=5$.) Note that diagrams involving $O_8$ have less than three
loops, because the definition of $O_8$ in \eq{OperBasis} involves one
power of the strong coupling $g_s$. We have also calculated the
contributions with penguin operators $O_{3-6}$ \cite{Asatrian:2017qaz},
counting their small Wilson coefficients as ${\cal O}(\alpha_s)$
\cite{NiersteNLO}, so that also here only one-loop and two-loop diagrams
are needed.

In the calculation one can neglect the charm mass on the lines attached
to a $O_{1,2}$ vertex, because the associated error is of order
$m_c^2/m_b^2$, i.e.\ 5\% of the expected ${\cal O} (15\%)$ NNLO
correction. A charm quark running in the closed quark loop in the
gluon propagator, however, leads to a term linear in $m_c/m_b$, so that
we have kept a non-zero charm mass there.   

For illustration we show the charm-loop contribution to the coefficient
multiplying $C_2^2$ in the NNLO correction to
$\widetilde G_S$:
\begin{align}
  F^{(2),N_V}_{S,22}(z) = &
                            - 9.01785 \log \frac{\mu_1}{m_b} - 11.8519 \log
                            \frac{\mu_2}{m_b} -14.2222 \log
                            \frac{\mu_1}{m_b} \log \frac{\mu_2}{m_b} +
                            10.6667 \log^2\frac{\mu_1}{m_b}
                            \nn 
  &   + 7.11111 \log^2\frac{\mu_2}{m_b} -
    42.0084 + 105.276 \frac{m_c}{m_b} +
    {\cal O} \left(\frac{m_c^2}{m_b^2} \right) \nonumber
\end{align}

Displaying only the error from the $\mu_1$ dependence, our NLO and
large-$N_f$ NNLO results read
\begin{align}
\dg^{NLO} &= \lt( 0.091 \pm 0.020_{\rm scale} \rt) \, \gev \qquad
\mbox{(pole)} \nn \dg^{NLO} &= \lt( 0.104 \pm 0.015_{\rm
scale}\rt)\, \gev \qquad \mbox{($\ov{\rm MS}$)} \label{eq:numres1}
\\[2mm]
\dg^{NNLO} &= \lt( 0.108 \pm 0.021_{\rm scale} \rt) \, \gev \qquad
\mbox{(pole)} \nn \dg^{NNLO} &= \lt( 0.103 \pm 0.015_{\rm scale}\rt)\,
\gev \qquad \mbox{($\ov{\rm MS}$)} \label{eq:numres2}
\end{align}
Note that we have used a different implementation of the $\ov{\rm MS}$
scheme here: In \eq{eq:nlo} the prefactor in \eq{eq:dghqe} is chosen
with $\bar m_b^2(\mu_1)$ and the $\mu_1$ dependence of this factor
nicely cancels with the one in $G^\prime$ and $\widetilde G_S$, and this
feature seems to be accidental. If one chooses $\bar m_b^2(\bar m_b)$
instead (with properly adjusted $\mu_1$ terms in
$G^\prime$ and $\widetilde G_S$), one finds the larger $\mu_1$
dependence of \eq{eq:numres1}. Our partial NNLO correction is sizable in
the pole scheme and lifts the result closer to the $\ov{\rm MS}$ result.
In the $\ov{\rm MS}$ scheme instead our large-$N_f$ correction is very
small and unlikely to be the dominant piece of the full NNLO result.

The large $N_f$ limit of QCD spoils asymptotic freedom, because the 
$\beta$ function changes sign for sufficiently large values of  $N_f$.
One may remedy this by ``naive non-Abelianisation (NNA)'', which means to
trade  $N_f$ for the leading coefficient 
$\beta_0=11-2/3 N_f$ of the QCD $\beta$ function
\cite{Brodsky:1982gc,Beneke:1994qe}. This procedure flips the sign of
the NNLO correction leading to
\begin{align}
\dg^{NNA} &= \lt( 0.071 \pm 0.020_{\rm scale} \rt) \, \gev \qquad
\mbox{(pole)} \nn
\dg^{NNA} &= \lt( 0.099 \pm 0.012_{\rm scale}\rt)\,
\gev \qquad \mbox{($\ov{\rm MS}$)}. \label{eq:nnanum}
\end{align}
In applications like ours, in which the size of the NNLO correction depends
on the chosen renormalisation scheme for the Wilson coefficients, it is
not clear whether NNA improves the result. E.g.\ in
Ref.~\cite{Asatrian:2010rq} it has been found that the
$\alpha_s^2 \beta_0$ term is not a good approximation to the full NNLO
result to the calculated quantity.

\section{Conclusions}
The calculation of the $\alpha_s^2 N_f$ terms of the NNLO correction to
\dg\ has reduced the renormalisation scheme dependence of the theory
prediction and has moved the pole scheme result close to the $\ov{\rm MS}$   
result. But there is no progress in the reduction of the dependence on the
renormalisation scale and the correction found in the $\ov{\rm MS}$
scheme is too small to be the dominant part of the full NNLO result.
Therefore a complete NNLO calculation is needed. In the meantime, we
advocate for the use of the $\ov{\rm MS}$ result with a conservative
perturbative error \cite{Asatrian:2017qaz}: 
\begin{align} 
\dg &= \lt( 0.104 \pm 0.015_{\rm scale} \pm
   0.007_{B,\widetilde B_S}
   \pm 0.015_{\lqcd/m_b} \rt)\, \gev
\qquad \mbox{($\ov{\rm MS}$)} . \label{eq:fin}
\end{align}

\Acknowledgements
I thank Hrachia Asatrian, Artyom Hovhannisyan, and Arsen Yeghiazaryan
for the fruitful collaboration on the project.  The presented work has
been supported by the German Bundesministerium f\"ur Bildung und
Forschung (BMBF) under contract no.~05H15VKKB1 and by VolkswagenStiftung
with grant no.~86426.

\end{document}